\documentclass[aps,prb,twocolumn,groupedaddress]{revtex4}
\usepackage{amsmath}
\usepackage{amssymb}
\usepackage{graphicx}

\begin{document}

\title{Thermally activated conductivity in gapped bilayer graphene}

\author{Maxim Trushin}

\affiliation{Institute for Theoretical Physics, University of Regensburg,
D-93040 Regensburg, Germany} 
\affiliation{Department of Physics, University of Texas, 1 University Station C1600, Austin, 78712 Texas, USA}

\begin{abstract}
This is a theoretical study of electron transport
in gated bilayer graphene --- a novel semiconducting material
with a tunable band gap.
It is shown that the {\em which-layer} pseudospin coherence
enhances the subgap conductivity
and facilitates the thermally activated transport.
The mechanism proposed can also lead to the non-monotonic
conductivity {\em vs.} temperature dependence at
a band gap size of the order of $10\,\mathrm{meV}$.
The effect can be observed in gapped bilayer graphene
sandwiched in boron nitride where the electron-hole
puddles and flexural phonons are strongly suppressed.
\end{abstract}

\maketitle

\section{Introduction}
\label{intro}

\begin{figure}
\centering\includegraphics[width=\columnwidth]{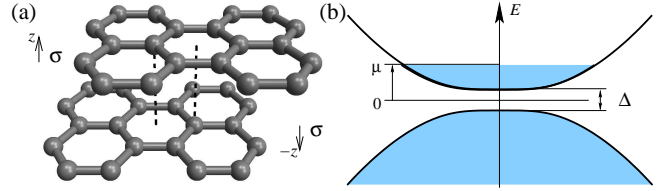}
\caption{Panel (a) shows bilayer graphene's crystal structure
and {\em which-layer} pseudospin orientation.
Panel (b) shows the lowest two bands and band gap size $\Delta$.
The chemical potential $\mu$ is counted from the middle of the band gap.}
\label{fig1}
\end{figure}

Graphene\cite{Nature2005novoselov} is often considered\cite{NatNan2010schwierz} as a 
most promising material for future
semiconductor industry. Indeed, it demonstrates high carrier mobility
even at room temperature\cite{Nature2007geim}
and is suitable for mass production
thanks to the chemical vapor deposition technique developed recently.
\cite{ACSNano2011jeon,NanoLett2009levendorf}
However, pristine graphene\cite{Nature2007geim} does not have a band gap which is
a crucial ingredient for the field effect transistor functionality.
It is possible to open the gap in {\em bilayer} graphene
by applying an external electric field perpendicular to the sample, see fig.~\ref{fig1}.
The effect was predicted by McCann \cite{PRB2006mccann}
and experimentally proven in ref.~\cite{PRL2007castro}.
Note that it is also possible to open a gap between hole and conduction
bands in bilayer graphene by means of an appropriate chemical doping.\cite{Science2006ohta}

In order to control the band gap and carrier density independently
the double-gated graphene devices have been utilized\cite{NatMat2008oostinga,PRL2010taychatanapat,PRB2010zou,Science2010weitz}.
The most striking feature observed is that the band gap obtained by infrared spectroscopy
\cite{Nature2009zhang,PRL2009mak} turns out to be
much too large to fit the thermally activated conductivity measurements.
There are a few attempts to resolve this discrepancy.
An earlier model\cite{PRL2007nilsson} suggests
the formation of midgap states in
which charge carriers are localized.
The band edge moves locally further into the gap
and a hopping mechanism dominates the conduction.\cite{NatMat2008oostinga,PRL2010taychatanapat}
The most recent approach\cite{PRL2011rossi}
employs fluctuations of the charged impurity potential
separating the electron and hole puddles. 
Indeed, the first experimental
observations\cite{NatMat2008oostinga,PRL2010taychatanapat,PRB2010zou}
of the insulating behavior in gapped bilayer graphene have
been made in the devices with graphene flakes placed
directly on the $\mathrm{SiO}_2$ substrate.
The substrate impurities are known to cause sizable potential fluctuations
which lead to the formation of electron-hole puddles
at low carrier densities.\cite{NatPhys2008martin}
If the substrate potential fluctuations are
strong enough then the small effective band gap
is expected to be due to the percolation through the charge inhomogeneities
overwhelming the real spectral gap.
The relevance of this mechanism to the subgap conductivity
is unquestionable as long as graphene
is placed on the $\mathrm{SiO}_2$ substrate.\cite{PRL2011rossi}
In recent experiments\cite{Science2010weitz}
carried out on {\em suspended} double-gated bilayer graphene
the electron-hole puddles are expected to be suppressed; nevertheless,
the activation energy deduced from the transport measurements
is still smaller than the band gap size.
An alternative model\cite{NatPhys2011li} suggests that the edge transport plays an important role in these measurements.\cite{Science2010weitz}
The phenomenon originates from non-trivial topological properties
of the electronic band structure in graphene which are similar to those
in spin-orbit induced topological insulators.\cite{RMP2010hasan} 

A question addressed in this paper is
whether there is another mechanism responsible for
the substantial subgap conductivity which can manifest
itself in gapped bilayer graphene sandwiched
in boron nitride.\cite{NanoLett2011mayorov,NatNano2010dean}
Such graphene samples are practically insusceptible to the environment
making the substrate much less important.
Moreover, the electron-hole puddles can be completely screened out
in double-layer systems similar to those recently reported 
in ref.~\cite{NatPhys2011ponomarenko}.
The edge transport, if any, can be precluded in Corbino geometry
which has been already utilized in recent experiments
carried out on double-gated bilayer graphene.\cite{NanoLett2010yan}
Using the pseudospin coherence concept 
we predict that the subgap conductivity contribution
does not vanish completely even though all abovementioned mechanisms are excluded,
see figs.~\ref{fig3},\ref{fig4}.
The signature of the mechanism in question is the non-monotonic
conductivity {\em vs.} temperature dependence at
a band gap size of a few tens of $\mathrm{meV}$, see figs.~\ref{fig5},\ref{fig6}.
This non-monotonic dependence could not be explained
within conventional model\cite{PRL2007nilsson,NatMat2008oostinga,PRL2010taychatanapat}
where disorder renormalizes the band gap to a smaller
value just by locally raising or lowering the band edges.

\begin{figure}
\centering\includegraphics[width=\columnwidth]{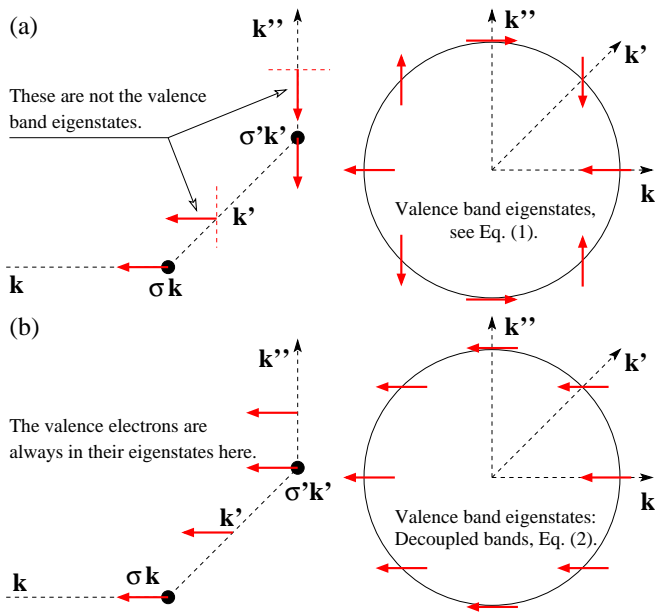}
\caption{Panel (a) schematically shows the in-plane pseudospin component $\propto\sin\vartheta_k$ for a valence band electron
while it is moving through disordered gapped bilayer graphene.
Each scattering event can be seen as
a classical ``measurement'' which changes the carrier's wave function
to a valence band eigenstate $\vert\sigma\mathbf{k}\rangle$ in accordance with the
momentum dependent pseudospin texture shown on the right.
At the same time the direction of particle's motion is changed
and since the older pseudospin orientation $\sigma$ does not correspond
to its eigenvalue with the new momentum $\mathbf{k'}$
the particle gets out of its eigenstate immediately after the scattering
event. The subsequent collision returns the particle to its eigenstates
with $\mathbf{k'}$ and $\sigma'$ but alters the direction of motion again.
Thus, the valence electron is obviously not in its equilibrium
valence eigenstate while it is moving between the scatterers and, therefore,
can make a contribution to the total conduction even
though the valence band is fully occupied.
Panel (b) corresponds to the valence band electrons described by
eq.~(\ref{hprim0}) where the pseudospin orientation does not depend
on the direction of motion. The particle remains in its eigenstate
after each scattering event. The valence band eigenstates can not conduct here as long
as the valence band is fully occupied.}
\label{fig2}
\end{figure}

\section{Concept}
\label{concept}

\begin{figure}
\centering\includegraphics[width=\columnwidth]{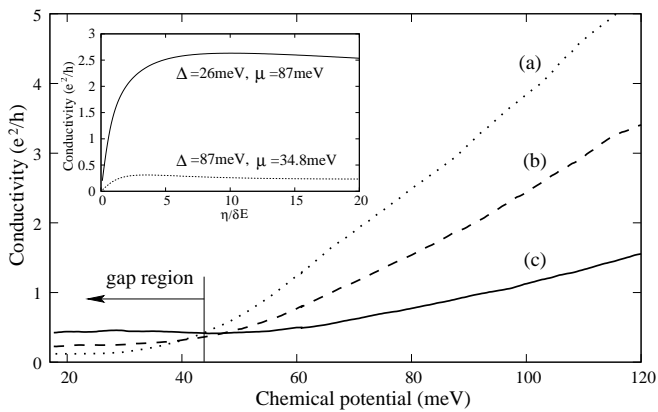}
\caption{Zero-temperature conductivity (for given spin and valley)
of gapped bilayer graphene (band gap size $\Delta=87\,\mathrm{meV}$)
in presence of the $\delta$-correlated
disorder with the strength $u_0=2.74\cdot 10^{-14}\,\mathrm{eVcm}^{2}$.
The concentration of scatterers $n_s=N_s/L^2$
(with $L=1.8\times 10^{-5}\,\mathrm{cm}$ being the sample size)
is different for each curve:
(a)  $0.54\cdot 10^{12}\,\mathrm{cm}^{-2}$,
(b)  $0.81\cdot 10^{12}\,\mathrm{cm}^{-2}$,
(c)  $1.62\cdot 10^{12}\,\mathrm{cm}^{-2}$.
The coupling $\eta$ is chosen to be equal to $10\delta E$, where
$\delta E =2\pi\hbar^2/L^2 m$.
The inset shows that the dependence of both metallic and subband conductivities
on $\eta$ is relatively weak in this case.}
\label{fig3}
\end{figure}

\begin{figure}
\centering\includegraphics[width=\columnwidth]{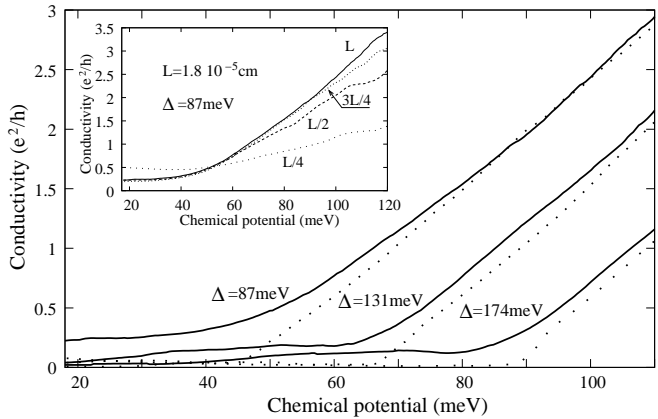}
\caption{Zero-temperature conductivities
for gapped bilayer graphene (solid lines) and
conventional intrinsic semiconductor (dotted lines)
described by eqs.~(\ref{h0}) and (\ref{hprim0}) respectively.
Note that the latter drops down to zero
as soon as the Fermi energy level reaches the bottom
of the conduction band. Disorder parameters are the same as
in fig.~\ref{fig3} for curve (b).
The inset shows how the conductivity curve changes under scaling.
The disorder concentration $n_s$ is chosen to be the same for all
$L$'s considered. One can see that the conductivity gets less sensitive to scaling
for larger $L$'s considered in this work.
}
\label{fig4}
\end{figure}

We argue that the difference between effective (transport) and 
actual (spectral) gaps is an intrinsic property of gapped bilayer graphene
following from the minimal two-band effective Hamiltonian
already employed in ref.~\cite{NatMat2008oostinga}.
The Hamiltonian can be written as $H_0=\vec{h}_k\cdot\vec{\sigma}$,
where
\begin{equation}
\label{h0}
\vec{h}_k=\frac{\hbar^2k^2}{2m}\left(\hat{x}\cos 2\varphi+\hat{y} \sin 2\varphi\right)
+\hat{z} U,
\end{equation}
and $\vec{\sigma}$ are Pauli matrices
representing the {\em pseudospin}\cite{NatPhys2006katsnelson} degree of freedom
for carriers in bilayer graphene which
originates from its peculiar crystal lattice
shown in fig.~\ref{fig1}(a)
with the $\sigma_z$-pseudospin projection referring to the layer index.
Here, $m$ is the effective mass, ${\mathbf k}$ is 
the two-component particle momentum, $\tan\varphi=k_y/k_x$,
and $\Delta=2U$ is the band gap.
The eigenvalues of $H_0$ are
$E_{\kappa\mathbf{k}}=\kappa\sqrt{\left(\frac{\hbar^2 k^2}{2m}\right)^2 + U^2}$
with $\kappa=\pm$ being the band index, and
the eigenstates are 
$\psi_{\kappa\mathbf{k}}(\mathbf{r})={\mathrm e}^{i\mathbf{kr}}|\chi_{\kappa k}\rangle$
with the spinors $|\chi_{+ k}\rangle=(\cos\frac{\vartheta_k}{2}, \sin\frac{\vartheta_k}{2}\mathrm{e}^{2i\varphi})^T$,
$|\chi_{- k}\rangle=(\sin\frac{\vartheta_k}{2}, -\cos\frac{\vartheta_k}{2}\mathrm{e}^{2i\varphi})^T$,
where $\cos\vartheta_k=U/\sqrt{\left(\frac{\hbar^2 k^2}{2m}\right)^2 + U^2}$.
The bands $E_{\kappa\mathbf{k}}$ are shown in fig.~\ref{fig1}(b).
In order to pinpoint the mechanism responsible for the transport gap renormalization
we compare the pseudospin-momentum coupled model 
(\ref{h0}) with the decoupled one in which $H'_0=\vec{h}'_k\cdot\vec{\sigma}$, where
\begin{equation}
\label{hprim0}
\vec{h}'_k=\hat{x}\frac{\hbar^2k^2}{2m}+\hat{z} U.
\end{equation}
The two models have the same energy spectrum but the
the eigenstate spinors do not depend on the direction of particle's motion here
and read  $|\chi'_{+ k}\rangle=(\cos\frac{\vartheta_k}{2}, \sin\frac{\vartheta_k}{2})^T$,
$|\chi'_{- k}\rangle=(\sin\frac{\vartheta_k}{2}, -\cos\frac{\vartheta_k}{2})^T$.
This only difference between two model Hamiltonians (\ref{h0})
and (\ref{hprim0}) leads to the drastic change in the subgap conductivity
behavior: The subgap conductivity of graphene does not vanish even
at zero temperature, see fig.~\ref{fig3},
whereas it does so within decoupled band model, as shown in fig.~\ref{fig4}.

The mechanism can be understood from fig.~\ref{fig2}.
Due to the pseudospin-momentum coupling in graphene 
the particle necessarily gets out of its valence band eigenstate
while moving between two subsequent collisions with the scatterers.
The resulting wave function, to a certain extent, can be seen
as a superposition between valence and conduction band states.
(The conduction band state obviously represents
an evanescent wave function as long as the energy
is below the bottom of conduction band.\cite{NatPhys2006katsnelson})
As consequence the electron and hole states become entangled
and in that way can facilitate the conductivity
making the effective band gap smaller than the actual one,
see figs.~\ref{fig3}--\ref{fig4}.

Note that such an entanglement has nothing to do with the electron-hole pairs.
The electron-hole pairs are entirely classical objects and occur in both graphene and
conventional semiconductor material as soon as the temperature
reaches the level high enough to excite the valence electrons across the band gap.
The interband entanglement is certainly of quantum mechanical nature.
This phenomenon, as many other effects related to the quantum mechanical coherence,
is sensitive to temperature.
In some cases one can observe the competition between the temperature-dependent
pseudospin decoherence and thermal activation of the electron-hole pairs
which results in the non-monotonic conductivity {\em vs.} temperature dependence,
see figs.~\ref{fig5}--\ref{fig6}.


\section{Methods} 
\label{meth}

To evaluate the {\em dc} conductivity $\sigma$ we follow the procedure described in \cite{PRL2007nomura,PRB2010trushin} and
start from the finite-size Kubo formula
\begin{equation}
\label{Kubo}
\sigma=-\frac{i\hbar e^2}{L^2}\sum\limits_{n,n'}
\frac{f^0_{E_n}-f^0_{E_{n'}}}{E_n - E_{n'}}\frac{\langle n \vert v_x \vert n'\rangle\langle n' \vert v_x \vert n\rangle}{E_n - E_{n'} + i\eta},
\end{equation}
where $L^2$ is the finite-size system area,
$\eta$ is the coupling to source and  drain reservoirs,
$\mathbf{v}$ is the velocity operator, $f^0_{E_n}$ is the
Fermi-Dirac distribution function, and $\vert n\rangle$ denotes an exact eigenstate
of the numerically solved Schr\"odinger equation 
for a finite-size disordered system with 
periodic boundary conditions: $(H_0+V)\psi_n= E_n \psi_n$,
where $V(\mathbf{r})=u_0\sum_i^{N_s} \delta(\mathbf{r}-\mathbf{R}_i)$
for the short-range disorder model we consider.
The scattering locations $\mathbf{R}_i$ and potential signs of $u_0$
are random. The Schr\"odinger equation has been solved using a 
large momentum-space cutoff $k^* \approx 7\cdot 10^6\,\mathrm{cm}^{-1}$
which corresponds to the energy scale at which the split-off bands of bilayer 
graphene become relevant and our two-band model is no longer applicable.\cite{PRB2010trushin}

The pseudospin-momentum coupling (the effect in which we are mainly interested here)
always occurs in graphene whichever disorder potential is assumed.
The model considered here should be seen as a generic one
where delta-correlated scattering potential is chosen
just for the sake of simplicity even though
the short range disorder mixes states in different valleys.
The intervalley scattering appears to be irrelevant to any other type of disorder
(charged impurities, ripples) and is therefore neglected here.
Note that the Thomas--Fermi theory has been 
recently employed\cite{PRL2011rossi}
to calculate the electronic structure in the
presence of the disorder potential due to charge impurities
in gapped bilayer graphene.
The theory is quasiclassical
and does not include the quantum mechanical entanglement considered here.
Most important is that
the amplitude of the screened disorder potential fluctuations must be of
the order of the gap size $\Delta$
in order to explain the difference between
the spectral band gap and the experimentally extracted transport gap.
Here, quite an opposite situation is considered:
The scatterer strength $u_0$ and concentration $n_s=N_s/L^2$
are chosen to be small enough ($u_0 n_s < \Delta$) to preclude
the percolative regime\cite{PRL2011rossi} and
substantial band gap renormalization.\cite{PRL2007nilsson} 
Such careful choice of disorder parameters 
makes it possible to observe the pseudospin coherence effects.

The Kubo conductivity (\ref{Kubo}) vanishes at $\eta\to 0$
as well as at $\eta\to \infty$. As one can see in fig.~\ref{fig3}(inset),
there is an intermediate region near $(\eta mL^2)/(2\pi\hbar^2)=10$
where the conductivity is not too sensitive to $\eta$.
It is natural to work in this region to estimate the conductivity
at a given system size $L$.
The length $L$ is chosen to be so large that
the conductivity curves don't change too much with further
increasing of $L$.
Fig.~\ref{fig4}(inset) shows the conductivity curves
for different sample sizes
starting from $\frac{1}{4}L=0.45\times 10^{-5}\,\mathrm{cm}$
with $0.45\times 10^{-5}\,\mathrm{cm}$ step.
One can see that the difference in the conductivity behavior
for the lengths $\frac{3}{4}L=1.35\times 10^{-5}\,\mathrm{cm}$
and $L=1.8\times 10^{-5}\,\mathrm{cm}$
becomes rather small, thus, the latter is chosen to be the typical
sample size which allows the scaling with $L$.
The typical scatterer number $N_s$ is a few hundreds for this $L$.
The momentum cut-off $k^*$ and $L$ fix the
Hamiltonian matrix dimension at $3362\times 3362$.

The effective mass $m$ for carriers in bilayer graphene
has been predicted to be equal to $0.054m_0$ \cite{SSC2007McCann} with
$m_0$ being the bare electron mass.
The latest measurements on suspended bilayer graphene \cite{Science2011mayorov}
indicate the effective mass of $0.028m_0$.
In contrast, the charge carriers in bilayer graphene on substrate demonstrate
larger effective mass about $0.04m_0$ \cite{PRB2011zou}
which turns out to be slightly different for electrons and holes.
Since the model considered here assumes the electron-hole symmetry
and the effects predicted below do not rely on the particular effective mass value 
such precision seems to be excessive for our purposes and
we choose just the round number $m=0.05m_0$.

The zero-temperature conductivity curves
depicted in figs.~\ref{fig3},\ref{fig4} are smoothed
by averaging over an energy interval
containing $10$--$100$ levels, over boundary conditions, 
and over several disorder potential realizations.\cite{PRB2010trushin}
The finite-temperature conductivity demonstrates much weaker
fluctuations, thus, the results shown in figs.~\ref{fig5},\ref{fig6}
are averaged just over a few disorder realizations.

\section{Results}
\label{res}

As one can see from fig.~\ref{fig3}, the conductivity
does not vanish even though the chemical potential $\mu$ gets
below the bottom of the conduction band and the temperature is zero.
Moreover, the subgap conductivity {\em increases}
with disorder (cf. ref.~\cite{PRL2007bardarson}). This peculiar behavior can be understood
in terms of the disorder-dependent quasiparticle life-time $\tau$ and
pseudospin decoherence time \cite{PRB2010trushin} $\tau_\mathrm{dc}=\hbar/2E_k$ 
with $E_k$ being the characteristic particle energy
which equals to either $\mu$ or $\Delta/2$ whichever is larger.
On the one hand, the interband entanglement is obviously weaker
for larger energies and stronger for smaller band gap sizes.
On the other hand, the evanescent components in the
interband entangled states become more important at shorter distances
and count in favor of strong disorder.
As consequence, the subgap pseudospin-coherent conductivity contribution 
increases with $\tau_\mathrm{dc}/\tau$ --- the effect we actually observe
in figs.~\ref{fig3},\ref{fig4}. The upper limit for quasiparticle life-time $\tau$ 
(which is the same as the momentum relaxation time in presence
of the short-range disorder potential)
can be estimated using the Fermi golden-rule at $\mu \gg U$ as 
$\tau\simeq 3\cdot10^{-14}\,\mathrm{s}$ corresponding
to the mobility $10^{3}\,\mathrm{cm}^{2}/\mathrm{Vs}$ for curve (b).

Looking at the plots in fig.~\ref{fig3} one might still think
that it is the impurity density of states, rather than the pseudospin-momentum
coupling, that is responsible for finite subgap conductivity.
In order to clarify this issue let us compare the pseudospin-momentum coupled model 
(\ref{h0}) with the decoupled one (\ref{hprim0}).
The two models have the same density of states but the
the eigenstate spinors do not depend on the direction of particle's motion in (\ref{hprim0}).
Here, either conduction or valence band eigenstate once created can propagate through
the disordered sample without changing its pseudospin orientation
even though the direction of motion is altered after each scattering event,
as shown in fig.~\ref{fig2}.
The interband entangled states do not occur here and
the conductivity vanishes as soon as the chemical potential
reaches the bottom of the conduction band, see dotted lines in fig.~\ref{fig4}.
In contrast, gapped bilayer graphene demonstrates
a substantial subgap conductivity at the same parameters.

Thus, to observe the substantial subgap conductivity
(i) the pseudospin must be coupled with
the particle momentum to create the interband entangled states
in disordered samples
and (ii) the system must be pseudospin-coherent, i. e. $\tau/\tau_\mathrm{dc}$ must be smaller than one.
Note that $\tau_\mathrm{dc}=\hbar/2\mu$ (for $\mu > \Delta/2$) decreases 
with increasing $\mu$
making the two conductivities in fig.~\ref{fig4}
indistinguishable at higher carrier concentrations.
On the other hand the quasiparticle life-time 
$\tau$ is longer in higher mobility samples with
less impurities and/or lighter carriers that requires
longer $\tau_\mathrm{dc}$ to fulfill the pseudospin coherence criteria.

\begin{figure}
\centering\includegraphics[width=\columnwidth]{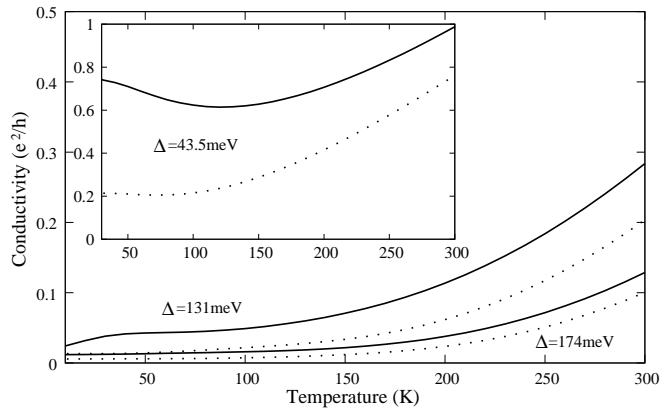}
\caption{Thermally activated conductivity at different band gap size $\Delta$
for gapped bilayer graphene (solid lines) and
decoupled band intrinsic semiconductor (dotted lines)
described by eqs.~(\ref{h0}) and (\ref{hprim0}) respectively.
The chemical potential is zero, i. e. it is placed exactly
in the middle of the gap.
The subgap conductivity increases slower with the
temperature within the decoupled band model.
The inset shows the competition between the temperature-dependent
pseudospin decoherence and thermal activation of the electron-hole pairs
resulting in the non-monotonic temperature dependence
of graphene's conductivity at smaller band gap.
Besides the band gap size shown in the plot,
all other parameters are the same as in fig.~\ref{fig3} for curve (b).
}
\label{fig5}
\end{figure}
\begin{figure}
\centering\includegraphics[width=\columnwidth]{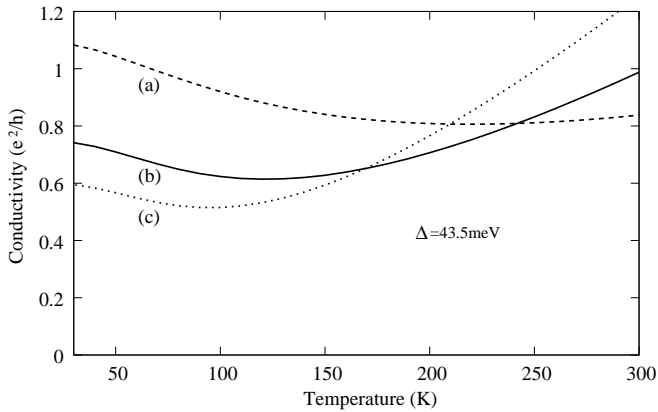}
\caption{This figure demonstrates
the non-monotonic behavior of thermally activated conductivity for bilayer graphene
at the intermediate band gap size $\Delta=43.5\,\mathrm{meV}$
for different disorder concentrations $n_s$.
The chemical potential is zero,
and disorder parameters for each curve are the same as in fig.~\ref{fig3}.
}
\label{fig6}
\end{figure}

The difference between bilayer graphene described by eq.~(\ref{h0})
and its rival with decoupled bands (\ref{hprim0})
at best can be seen in the thermally activated conductivity.
The calculations can also be considered as a simulation of the charge transport
in a field effect transistor turned to the ``off'' state
when the chemical potential is placed exactly in the middle
of the band gap hampering both electron and hole transport at low temperatures.
As one can see in  fig.~\ref{fig5}, the pseudospin-coupled carriers can be
excited easier than the decoupled ones.
The difference between conductivities in these two cases
becomes essential at room temperatures.
Note that if $T\ll \Delta$, then the {\em pseudospin-incoherent} 
conductivity can be well described by the classical formula
$\frac{\tau T}{\hbar}\exp(-\frac{\Delta}{2T})$
indicating that the thermally activated conductivity always
increases with temperature.
In contrast, the subgap {\em pseudospin-coherent} conductivity decreases
as soon as $T$ becomes comparable with $\Delta/2$
substituting the latter in the expression for $\tau_\mathrm{dc}$
and breaking down the pseudospin-coherence.
The competition between these two mechanisms can
result in the non-monotonic temperature dependence
of graphene's conductivity, see fig.~\ref{fig5}(inset).
Note that if $T\gg \Delta$, then both conductivity curves coincide.
(This regime is not shown in figure.)

The non-monotonic conductivity behavior is robust under
moderate change of the disorder strength, see fig.~\ref{fig6}.
However, as it was mentioned before,
the disorder strength $n_s u_0$ must always be smaller than the band gap size
in order to preclude the influence of midgap states.
The bilayer samples must therefore be relatively clean to observe
the non-monotonic conductivity behavior predicted here.
The necessary quality can probably be
achieved in graphene on boron nitride.\cite{NatNano2010dean}
It is also important that the phonons, which are not
considered here at all, might spoil the effect.
The phonon resistivity contribution in bilayer graphene
is dominated by flexural phonons and rapidly increases with temperature.\cite{PRB2011ochoa}
The flexural phonons can be again suppressed in graphene sandwiched between
boron nitride layers.\cite{NanoLett2011mayorov,NatNano2010dean,NatPhys2011ponomarenko}

\section{Conclusion}
\label{concl}

To conclude, there is a fundamental obstacle
which limits the functionality of the field effect transistor
based on gapped bilayer graphene.
The physical mechanism responsible for that is intimately linked
to the pseudospin-momentum coupling which leads to
the instantaneous generation of the interband entangled states
in the presence of disorder.
It makes higher ``leakage'' current in the ``off''
state and therefore limits the possible on/off ratio by lower values
as compared to those in conventional semiconductor devices
with the same mobility and band gap size.
In contrast to the ``leakage'' mechanisms considered before, \cite{PRL2007nilsson,NatPhys2011li,PRL2011rossi}
the interband entanglement described here is unavoidable unless the very
crystal lattice is broken.
Moreover, in contrast to the universal subgap conductivity observed
in the topological insulators, \cite{RMP2010hasan} the subgap 
conductivity in bilayer graphene turns out to be sensitive
to the band gap size and disorder strength.
The non-monotonic conductivity {\em vs.} temperature dependence
predicted here can be seen as a signature of the pseudospin precession
responsible for the difference between the transport and spectral gaps.
The effect can probably be observed in doubly gated 
bilayer graphene sandwiched between boron nitride layers where
the charge inhomogeneity and flexural phonon conductivity contributions
are substantially reduced.

\begin{acknowledgments}
I would like to thank Prof. Allan MacDonald for his hospitality
during my stay at the University of Texas at Austin, where
a part of this work has been done,
and DFG for financial support through the project TR 1019/1-1.
\end{acknowledgments}

\bibliography{graphene.bib}

\end{document}